# Online Learning and Resource-Bounded Dimension:
## Winnow Yields New Lower Bounds for Hard Sets*

John M. Hitchcock
Department of Computer Science

**Abstract**

We establish a relationship between the online mistake-bound model of learning and resource-bounded dimension. This connection is combined with the Winnow algorithm to obtain new results about the density of hard sets under adaptive reductions. This improves previous work of Fu (1995) and Lutz and Zhao (2000), and solves one of Lutz and Mayordomo's "Twelve Problems in Resource-Bounded Measure" (1999).

## 1 Introduction

This paper has two main contributions: (i) establishing a close relationship between resource-bounded dimension and Littlestone's online mistake-bound model of learning, and (ii) using this relationship along with the Winnow algorithm to resolve an open problem in computational complexity. In this introduction we briefly describe these contributions.

### 1.1 Online Learning and Dimension

Lindner, Schuler, and Watanabe [15] studied connections between computational learning theory and resource-bounded measure, primarily working with the probably approximately correct (PAC) model. They also included the observation that any "admissible" subclass of P/poly that is polynomial-time learnable in Angluin's exact learning model [2] must have p-measure 0. The proof of this made use of the essential equivalence between Angluin's model and Littlestone's online mistake-bound model [16].

In the online mistake-bound model, a learner is presented a sequence of examples, and is asked to predict whether or not they belong to some unknown target concept. The concept is drawn from some concept class, which is known to the learner, and the examples may be chosen by an adversary. After making its prediction about each example, the learner is told the correct classification for the example, and learner may use this knowledge in making future predictions. The mistake bound of the learner is the maximum number of incorrect predictions the learner will make, over any choice of target concept and sequence of examples.

We push the observation of [15] much further, developing a powerful, general framework for showing that classes have resource-bounded *dimension* 0. Resource-bounded measure and dimension involve betting on the membership of strings in an unknown set. To prove that a class has

*This research was supported in part by National Science Foundation grant 0515313.

dimension 0, we show that it suffices to give a reduction to a family of concept classes that has a good mistake-bound learning algorithm. It is possible that the reduction can take exponential-time and that the learning algorithm can also take exponential-time, as long as the mistake bound of the algorithm is subexponential. If we have a reduction from the unknown set to a concept in learnable concept class, we can view the reduction as generating a sequence of examples, apply the learning algorithm to these examples, and use the learning algorithm's predictions to design a good betting strategy. Formal details of this framework are given in Section 3.

## 1.2 Density of Hard Sets

The two most common notions of polynomial-time reductions are many-one ($\leq_m^P$) and Turing ($\leq_T^P$). A many-one reduction from $A$ to $B$ maps instances of $A$ to instance of $B$, preserving membership. A Turing reduction from $A$ to $B$ makes many, possibly adaptive, queries to $B$ in order to solve $A$. Many-one reductions are a special case of Turing reductions. In between $\leq_m^P$ and $\leq_T^P$ is a wide variety of polynomial-time reductions of different strengths.

A common use of reductions is to demonstrate hardness for a complexity class. Let $\leq_\tau^P$ be a polynomial-time reducibility. For any set $B$, let $P_\tau(B) = \{A \mid A \leq_\tau^P B\}$ be the class of all problems that $\leq_\tau^P$-reduce to $B$. We say that $B$ is $\leq_\tau^P$-*hard* for a complexity class $\mathcal{C}$ if $\mathcal{C} \subseteq P_\tau(B)$, that is, every problem in $\mathcal{C}$ $\leq_\tau^P$-reduces to $B$. For a class $\mathcal{D}$ of sets, a useful notation is $P_\tau(\mathcal{D}) = \bigcup_{B \in \mathcal{D}} P_\tau(B)$.

A problem $B$ is *dense* if there exists $\epsilon > 0$ such that $|B_{\leq n}| > 2^{n^\epsilon}$ for all but finitely many $n$. All known hard sets for the exponential-time complexity classes $E = \text{DTIME}(2^{O(n)})$ or $\text{EXP} = \text{DTIME}(2^{n^{O(1)}})$ are dense. Whether every hard set must be dense has been often studied. First, Meyer [25] showed that every $\leq_m^P$-hard set for E must be dense, and he observed that proving the same for $\leq_T^P$-reductions would imply that E has exponential circuit-size complexity. Since then, a line of research has obtained results for a variety of reductions between $\leq_m^P$ and $\leq_T^P$, specifically the conjunctive ($\leq_c^P$) and disjunctive ($\leq_d^P$) reductions, and for various functions $f(n)$, the bounded query $\leq_{f(n)-\text{tt}}^P$ and $\leq_{f(n)-T}^P$ reductions:

1. Watanabe [27, 10] showed that every hard set for E under the $\leq_c^P$, $\leq_d^P$, or $\leq_{O(\log n)-\text{tt}}^P$ reductions is dense.

2. Lutz and Mayordomo [20] showed that for all $\alpha < 1$, the class $P_{n^\alpha-\text{tt}}(\text{DENSE}^c)$ has p-measure 0, where DENSE is the class of all dense sets. Since E does not have p-measure 0, their result implies that every $\leq_{n^\alpha-\text{tt}}^P$-hard set for E is dense.

3. Fu [8] showed that for all $\alpha < 1/2$, every $\leq_{n^\alpha-T}^P$-hard set for E is dense, and that for all $\alpha < 1$, every $\leq_{n^\alpha-T}^P$-hard set for EXP is dense.

4. Lutz and Zhao [22] gave a measure-theoretic strengthening of Fu's results, showing that for all $\alpha < 1/2$, $P_{n^\alpha-T}(\text{DENSE}^c)$ has p-measure 0, and that for all $\alpha < 1$, $P_{n^\alpha-T}(\text{DENSE}^c)$ has $p_2$-measure 0.

This contrast between E and EXP in the last two references was left as a curious open problem, and exposited by Lutz and Mayordomo [21] as one of their "Twelve Problems in Resource-Bounded Measure":

> **Problem 6.** *For $\alpha \leq \frac{1}{2} < 1$, is it the case that $P_{n^\alpha-T}(\text{DENSE}^c)$ has p-measure 0 (or at least, that $E \not\subseteq P_{n^\alpha-T}(\text{SPARSE})$)?*

We resolve this problem, showing the much stronger conclusion that the classes in question have p-dimension 0. But first, in Section 4, we prove a theorem about disjunctive reductions that illustrates the basic idea of our technique. We show that the class $P_d(\text{DENSE}^c)$ has p-dimension 0. The proof uses the learning framework of Section 3 and Littlestone's Winnow algorithm [16]. Suppose that $A \leq_d^p S$, where $S$ is a nondense set. Then there is a reduction $g$ mapping strings to sets of strings such that $x \in A$ if and only if at least one string in $g(x)$ belongs to $S$. We view the reduction $g$ as generating examples that we can use to learn a disjunction based on $S$. Because $S$ is subexponentially dense, the target disjunction involves a subexponential number of variables out of exponentially many variables. This is truly a case "when irrelevant attributes abound" [16] and the Winnow algorithm perfoms exceedingly well to establish our dimension result. In the same section we also use the learning framework to show that $P_c(\text{DENSE}^c)$ has p-dimension 0. These results give new proofs of Watanabe's aforementioned theorems about $\leq_d^p$-hard and $\leq_c^p$-hard sets for E.

Our main theorem, presented in Section 5, is that for all $\alpha < 1$, $P_{n^\alpha-T}(\text{DENSE}^c)$ has p-dimension 0. This substantially improves the results of [20, 8, 22]. The resource-bounded measure proofs in [20, 22] use the concept of *weak stochasticity*. As observed by Mayordomo [24], this stochasticity approach can be extended to show a $-1^{\text{st}}$-*order scaled dimension* [12] result, but it seems a different technique is needed for an (unscaled) dimension result. Our learning framework turns out to be just what is needed. We reduce the class $P_{n^\alpha-T}(\text{DENSE}^c)$ to a family of learnable disjunctions. For this, we make use of a technique that Allender, Hemaspaandra, Ogiwara, and Watanabe [1] used to prove a surprising result converting bounded-query reductions to sparse sets into disjunctive reductions to sparse sets: $P_{\text{btt}}(\text{SPARSE}) \subseteq P_d(\text{SPARSE})$. Carefully applying the same technique on a sublinear-query Turing-reduction to a nondense set results in a disjunction with a nearly exponential blowup, but it can still be learned by Winnow in our dimension setting.

The density of complete and hard sets for NP has also been studied often, with motivation coming originally from the Berman-Hartmanis isomorphism conjecture [5]: all many-one complete sets are dense if the isomorphism conjecture holds. Since no absolute results about the density of NP-complete or NP-hard sets can be proved without separating P from NP, the approach has been to prove conditional results under a hypothesis on NP. Mahaney [23] showed that if $P \neq NP$, then no sparse set is $\leq_m^P$-hard for NP. Ogiwara and Watanabe [26] extended Mahaney's theorem to the $\leq_{\text{btt}}^P$-hard sets. Deriving a result from $P \neq NP$ about NP-hard sets under unbounded truth-table reductions is still an open problem, but a measure-theoretic assumption yields very strong consequences. Lutz and Zhao [22] showed that under the hypothesis "NP does not have p-measure 0," every $\leq_{n^\alpha-T}^P$-hard set for NP must be dense, for all $\alpha < 1$. In Section 6 we present the same conclusion under the weaker hypothesis "NP has positive p-dimension," and some additional consequences.

## 2 Preliminaries

The set of all binary strings is $\{0,1\}^*$. The length of a string $x \in \{0,1\}^*$ is $|x|$. We write $\lambda$ for the empty string. For $n \in \mathbb{N}$, $\{0,1\}^n$ is the set of strings of length $n$ and $\{0,1\}^{\leq n}$ is the set of strings of length at most $n$.

A *language* is a subset $L \subseteq \{0,1\}^*$. We write $L_{\leq n} = L \cap \{0,1\}^{\leq n}$ and $L_{=n} = L \cap \{0,1\}^n$. We say that $L$ is *sparse* if there is a polynomial $p(n)$ such that for all $n$, $|L_{=n}| \leq p(n)$. We say that $L$ is *(exponentially) dense* if there is a constant $\epsilon > 0$ such that $|L_{\leq n}| > 2^{n^\epsilon}$ for all sufficiently large $n$. We write SPARSE and DENSE for the classes of all sparse languages and all dense languages. The complement DENSE$^c$ of DENSE is the class of all *nondense* languages.

## 2.1 Polynomial-Time Reductions

We use standard notions of polynomial-time reducibilities:

- *Turing reducibility:* $A \leq^\mathrm{P}_\mathrm{T} B$ if there is a polynomial-time oracle machine $M$ such that $A = L(M^B)$.

- *Truth-table reducibility:* $A \leq^\mathrm{P}_\mathrm{tt} B$ if there is a polynomial-time oracle machine $M$ that makes nonadaptive queries such that $A = L(M^B)$.

- *Disjunctive reducibility:* $A \leq^\mathrm{P}_\mathrm{d} B$ if there is a polynomial-time computable $f : \{0,1\}^* \to \mathcal{P}(\{0,1\}^*)$ such that for all $x$, $x \in A$ if and only if $f(x) \cap B \neq \emptyset$.

- *Conjunctive reducibility:* $A \leq^\mathrm{P}_\mathrm{c} B$ if there is a polynomial-time computable $f : \{0,1\}^* \to \mathcal{P}(\{0,1\}^*)$ such that for all $x$, $x \in A$ if and only if $f(x) \subseteq B$.

We write $\leq^\mathrm{P}_{q(n)-\mathrm{T}}$ or $\leq^\mathrm{P}_{q(n)-\mathrm{tt}}$ to indicate that the reduction makes at most $q(n)$ queries on any input of length $n$. The bounded reducibility $A \leq^\mathrm{P}_\mathrm{btt} B$ means $A \leq^\mathrm{P}_{k-\mathrm{tt}} B$ for some constant $k$.

Let $\leq^\mathrm{P}_\tau$ be a polynomial-time reducibility. For any language $B$, we define $\mathrm{P}_\tau(B) = \{A \mid A \leq^\mathrm{P}_\tau B\}$. A language $B$ is $\leq^\mathrm{P}_\tau$-*hard* for a class $\mathcal{C}$ if $\mathcal{C} \subseteq \mathrm{P}_\tau(B)$. For any class $\mathcal{D}$ of languages, $\mathrm{P}_\tau(\mathcal{D}) = \bigcup_{B \in \mathcal{D}} \mathrm{P}_\tau(B)$.

## 2.2 Resource-Bounded Measure and Dimension

Resource-bounded measure and dimension were introduced in [17, 19, 4]. Here we briefly review the definitions and basic properties. We refer to the original sources and also the surveys [18, 21, 13] for more information.

The *Cantor space* is $\mathbf{C} = \{0,1\}^\infty$. Each language $A \subseteq \{0,1\}^*$ is identified with its characteristic sequence $\chi_A \in \mathbf{C}$ according to the standard (lexicographic) enumeration of $\{0,1\}^*$. We typically write $A$ in place of $\chi_A$. In this way a complexity class $\mathcal{C} \subseteq \mathcal{P}(\{0,1\}^*)$ is viewed as a subset $\mathcal{C} \subseteq \mathbf{C}$. We use the notation $S \restriction n$ to denote the first $n$ bits of a sequence $S \in \mathbf{C}$.

Let $s > 0$ be a real number. An *$s$-gale* is a function $d : \{0,1\}^* \to [0, \infty)$ such that for all $w \in \{0,1\}^*$,
$$d(w) = \frac{d(w0) + d(w1)}{2^s}.$$
A *martingale* is a 1-gale.

The goal of an $s$-gale is to obtain large values on sequences:

**Definition.** Let $d$ be an $s$-gale and $S \in \mathbf{C}$.

1. $d$ *succeeds on* $S$ if $\limsup_{n \to \infty} d(S \restriction n) = \infty$.

2. $d$ *succeeds strongly on* $S$ if $\liminf_{n \to \infty} d(S \restriction n) = \infty$.

3. The *success set* of $d$ is $S^\infty[d] = \{S \in \mathbf{C} \mid d \text{ succeeds on } S\}$.

4. The *strong success set* of $d$ is $S_{\text{str}}^\infty[d] = \{S \in \mathbf{C} \mid d \text{ succeeds strongly on } S\}$.

Notice that the smaller $s$ is, the more difficult it is for an $s$-gale to obtain large values. Succeeding martingales ($s = 1$) imply measure 0, and the infimum $s$ for which an $s$-gale can succeed (or strongly succeed) gives the dimension (or strong dimension):

**Definition.** Let $X \subseteq \mathbf{C}$.

1. $X$ has p-*measure 0*, written $\mu_{\text{p}}(X) = 0$, if there is a polynomial-time computable martingale $d$ such that $X \subseteq S^\infty[d]$.

2. The p-*dimension* of $X$, written $\dim_{\text{p}}(X)$, is the infimum of all $s$ such that there exists a polynomial-time computable $s$-gale $d$ with $X \subseteq S^\infty[d]$.

3. The *strong* p-*dimension* of $X$, written $\text{Dim}_{\text{p}}(X)$, is the infimum of all $s$ such that there exists a polynomial-time computable $s$-gale $d$ with $X \subseteq S_{\text{str}}^\infty[d]$.

We now summarize some of the basic properties of the p-dimensions and p-measure.

**Proposition 2.1.** ([19, 4]) *Let $X, Y \subseteq \mathbf{C}$.*

1. $0 \leq \dim_{\text{p}}(X) \leq \text{Dim}_{\text{p}}(X) \leq 1$.

2. *If* $\dim_{\text{p}}(X) < 1$, *then* $\mu_{\text{p}}(X) = 0$.

3. *If* $X \subseteq Y$, *then* $\dim_{\text{p}}(X) \leq \dim_{\text{p}}(Y)$ *and* $\text{Dim}_{\text{p}}(X) \leq \text{Dim}_{\text{p}}(Y)$.

The following theorem indicates that the p-dimensions are useful for studies within the complexity class E.

**Theorem 2.2.** ([17, 19, 4])

1. $\mu_{\text{p}}(\text{E}) \neq 0$. *In particular,* $\dim_{\text{p}}(\text{E}) = \text{Dim}_{\text{p}}(\text{E}) = 1$.

2. *For all* $c \in \mathbb{N}$, $\text{Dim}_{\text{p}}(\text{DTIME}(2^{cn})) = 0$.

## 2.3 Online Mistake-Bound Model of Learning

A *concept* is a set $C \subseteq U$, where $U$ is some *universe*. A concept $C$ is often identified with its characteristic function $f_C : U \to \{0, 1\}$. In this paper the universe is always a set of binary strings. A *concept class* is a set of concepts $\mathcal{C} \subseteq \mathcal{P}(U)$.

Given a concept class $\mathcal{C}$ and a universe $U$, a learning algorithm tries to learn an unknown target concept $C \in \mathcal{C}$. The algorithm is given a sequence of examples $x_1$, $x_2$, ... in $U$. When given each example $x_i$, the algorithm must predict if $x_i \in C$ or $x_i \notin C$. The algorithm is then told the correct answer and given the next example. The algorithm makes a *mistake* if its prediction for membership of $x_i$ in $C$ is wrong. This proceeds until every member of $U$ is given as an example.

The goal is to minimize the number of mistakes. The *mistake bound* of a learning algorithm $A$ for a concept class $\mathcal{C}$ is the maximum over all $C \in \mathcal{C}$ of the number of mistakes $A$ makes when learning $C$, over all possible sequences of examples. The *running time* of $A$ on $\mathcal{C}$ is the maximum time $A$ takes to make a prediction.

## 2.4 Disjunctions and Winnow

An interesting concept class is the class of monotone disjunctions, which can be efficiently learned by Littlestone's Winnow algorithm [16]. A monotone disjunction on $\{0,1\}^n$ is a formula of the form $\phi_V = \bigvee_{i \in V} x_i$, where $V \subseteq \{1, \ldots, n\}$ and we write a string $x \in \{0,1\}^n$ as $x = x_1 \cdots x_n$. The concept $\phi_V$ can also be viewed as the set $\{x \in \{0,1\}^n \mid \phi_V(x) = 1\}$ or equivalently as $\{A \subseteq \{1, \ldots, n\} \mid A \cap V \neq \emptyset\}$.

The Winnow algorithm has two parameters $\alpha$ (a weight update multiplier) and $\theta$ (a threshold value). Initially, each variable $x_i$ has a weight $w_i = 1$. To classify a string $x$, the algorithm predicts that $x$ is in the concept if $\sum_i w_i x_i > \theta$, and not in the concept otherwise. The weights are updated as follows whenever a mistake is made.

- If a negative example $x$ is incorrectly classified, then set $w_i := 0$ for all $i$ such that $x_i = 1$. (Certainly these $x_i$'s are not in the disjunction.)

- If a positive example $x$ is incorrectly classified, then set $w_i := \alpha \cdot w_i$ for all $i$ such that $x_i = 1$. (It is considered more likely that these $x_i$'s are in the disjunction.)

A useful setting of the parameters is $\alpha = 2$ and $\theta = n/2$. With these parameters, Littlestone proved that Winnow will make at most $2k \log n + 2$ mistakes when the target disjunction has at most $k$ literals. Also, the algorithm uses $O(n)$ time to classify each example and update the weights.

## 3 Learning and Dimension

In this section we present a framework relating online learning to resource-bounded dimension. This framework is based on reducibility to learnable concept class families.

**Definition.** A sequence $\mathcal{C} = (\mathcal{C}_n \mid n \in \mathbb{N})$ of concept classes is called a *concept class family*.

We consider two types of reductions:

**Definition.** Let $L \subseteq \{0,1\}^*$, $\mathcal{C} = (\mathcal{C}_n \mid n \in \mathbb{N})$ be a concept class family, and $r(n)$ be a time bound.

1. We say $L$ *strongly reduces to* $\mathcal{C}$ *in* $r(n)$ *time*, and we write $L \leq^r_{str} \mathcal{C}$, if there exists a sequence of target concepts $(c_n \in \mathcal{C}_n \mid n \in \mathbb{N})$ and a reduction $f$ computable in $O(r(n))$ time such that for all but finitely many $n$, for all $x \in \{0,1\}^n$, $x \in L$ if and only if $f(x) \in c_n$.

2. We say $L$ *weakly reduces to* $\mathcal{C}$ *in* $r(n)$ *time*, we write $L \leq^r_{wk} \mathcal{C}$ if there a reduction $f$ computable in $O(r(n))$ time such that for infinitely many $n$, there is a concept $c_n \in \mathcal{C}_n$ such that for all $x \in \{0,1\}^{\leq n}$, $x \in L$ if and only if $f(0^n, x) \in c_n$.

It is necessary to quantify both the time complexity and mistake bound for learning a concept class family:

**Definition.** Let $t, m : \mathbb{N} \to \mathbb{N}$ and let $\mathcal{C} = (\mathcal{C}_n \mid n \in \mathbb{N})$ be a concept class family. We say that $\mathcal{C} \in \mathcal{L}(t, m)$ if there is an algorithm that learns $\mathcal{C}_n$ in $O(t(n))$ time with mistake bound $m(n)$.

Combining the two previous definitions we arrive at our central technical concept:

**Definition.** Let $r, t, m : \mathbb{N} \to \mathbb{N}$.

1. $\mathcal{RL}_{\text{str}}(r, t, m)$ is the class of all languages that $\leq^r_{str}$-reduce to some concept class family in $\mathcal{L}(t, m)$.

2. $\mathcal{RL}_{\text{wk}}(r, t, m)$ is the class of all languages that $\leq^r_{wk}$-reduce to some concept class family in $\mathcal{L}(t, m)$.

A remark about the parameters in this definition is in order. If $A \in \mathcal{RL}_{\text{str}}(r, t, m)$, then $A \leq^r_{str} \mathcal{C}$ for some concept class family $\mathcal{C} = (\mathcal{C}_n \mid n \in \mathbb{N})$. Then $x \in A_{=n}$ if and only if $f(x) \in c_n$, where $c_n \in \mathcal{C}_n$ is the target concept and $f$ is the reduction. We emphasize that the complexity of learning $\mathcal{C}_n$ is measured in terms of $n = |x|$, and not the size of $c_n$ or $f(x)$. Instead $\mathcal{C}_n$ is learnable in time $O(t(n))$ with mistake bound $m(n)$.

The following theorem is the main technical tool in this paper. Here we consider exponential-time reductions to concept classes that can be learned in exponetial-time, but with subexponentially-many mistakes.

**Theorem 3.1.** *Let $c \in \mathbb{N}$.*

1. $\mathcal{RL}_{\text{str}}(2^{cn}, 2^{cn}, o(2^n))$ *has strong* p-*dimension 0.*

2. $\mathcal{RL}_{\text{wk}}(2^{cn}, 2^{cn}, o(2^n))$ *has* p-*dimension 0.*

*Proof.* We only prove that $\mathcal{RL}_{\text{wk}}(2^{cn}, 2^{cn}, o(2^n))$ has p-dimension 0. The other part of the theorem is proved similarly. Let $s > 0$ such that $2^s$ is rational. It suffices to show that the class has p-dimension at most $s$.

Let $A \in \mathcal{RL}_{\text{wk}}(2^{cn}, 2^{cn}, o(2^n))$. Then there is a concept class family $\mathcal{C} = \{\mathcal{C}_n \mid n \in \mathbb{N}\} \in \mathcal{L}(2^{cn}, o(2^n))$ such that $A \leq^{2^{cn}}_{wk} \mathcal{C}$. Let $f$ be this reduction from $A$ to $\mathcal{C}$. The for infinitely many $n$, there is a target concept $c_n \in \mathcal{C}_n$ such that

$$x \in A_{\leq n} \iff f(x) \in c_n.$$

Let $J$ be the set of all $n$ such that this concept exists. Let $\mathcal{A}$ be a $2^{cn}$-time learning algorithm for $\mathcal{C}$ with mistake bound $o(2^n)$.

Fix an $n$ and let $N = 2^{n+1} - 1$. We view the reduction $f$ as generating a sequence of examples

$$f(s_0), f(s_1), \ldots, f(s_N),$$

one for each string in $\{0, 1\}^{\leq n}$. The idea is to run the algorithm $\mathcal{A}$ on this sequence of examples, trying to learn $c_n$. We will use $\mathcal{A}$'s predictions to define an $s$-gale $d_n$ inductively as follows.

1. Let $N_0 = 2^{n/2}$. For all strings $w$ with $|w| < N_0$, $d_n(w) = 2^{(s-1)|w|}$.

2. Let $\epsilon$ be a small rational number to be determined later. Let $w$ be any string $w$ with $N_0 \leq |w| < N$. Run $\mathcal{A}$ on the sequence of examples $f(s_{N_0}), \ldots, f(s_{|w|})$, telling $\mathcal{A}$ that for each $i$, $N_0 \leq i < |w|$,

    - If $w[i] = 1$, then $f(s_i)$ is a positive example.
    - If $w[i] = 0$, then $f(s_i)$ is a negative example.

    At the end $\mathcal{A}$ will output a prediction for $f(s_{|w|})$.

- If $\mathcal{A}$ predicts that $f(s_{|w|})$ is a member of the target concept $c_n$, then we let
  - $d_n(w1) = 2^s(1-\epsilon)d(w)$,
  - $d_n(w0) = 2^s\epsilon d(w)$.
- Otherwise, $\mathcal{A}$ predicts that $f(s_{|w|})$ is not a member of the target concept $c_n$, and we let
  - $d_n(w0) = 2^s(1-\epsilon)d(w)$,
  - $d_n(w1) = 2^s\epsilon d(w)$.

3. For $w$ with $|w| \geq N$, we set $d_n(w0) = d_n(w1) = 2^{(s-1)}d_n(w)$.

The reason for making $d_n$ wait until $N_0$ to bet is computational efficiency. For $|w| < N_0$, $d_n(w)$ is computable in $O(|w|)$ time. If $|w| \geq N_0$, then to compute $d_n(w)$ we need to execute $\mathcal{A}$ on at most $|w|$ examples, each execution taking $O(2^{cn})$ time to compute the example and $O(2^{cn})$ to compute the label, for a total time of $O(|w|2^{cn})$. Because $|w| \geq 2^{n/2}$, this simplifies to $O(|w|^{2c+1})$.

Each time $\mathcal{A}$ makes a correct prediction, the value of the $s$-gale is increased by a $2^s(1-\epsilon)$ factor. When $\mathcal{A}$ makes a mistake, the value is multiplied by $2^s\epsilon$. Let $w_n$ be the length $N$ prefix of $A$'s characteristic sequence and suppose that $n \in J$. In the computation of $d_n(w_n)$, observe that $\mathcal{A}$ is told the correct labels for the examples according to the target concept $c_n$. Let $m_n$ be the number of mistakes that $\mathcal{A}$ makes on this sequence of examples when learning $c_n$; we know that $m_n = o(2^n)$. Then

$$\begin{aligned}
d(w_n) &= 2^{s(N-N_0)} \cdot (1-\epsilon)^{N-N_0-m_n} \cdot \epsilon^{m_n} \cdot 2^{(s-1)N_0} \\
&= 2^{sN+[(N-N_0-m_n)\log(1-\epsilon)]+[m_n \log \epsilon]-N_0} \\
&\geq 2^{sN-\left(N\log\frac{1}{1-\epsilon}+m_n \log\frac{1-\epsilon}{\epsilon}\right)-N_0}.
\end{aligned}$$

We choose $\epsilon \in \mathbb{Q}$ so that $\log\frac{1}{1-\epsilon} < s$ and let $0 < \delta < s - \frac{1}{1-\epsilon}$. Then since $m_n$ and $N_0$ are both $o(N)$, when $n \in J$ is large enough we have

$$d_n(w_n) \geq 2^{\delta N}.$$

Let $d$ be the $s$-gale $d = \sum_{n=1}^{\infty} 2^{-n}d_n$. Then $A \in S^{\infty}[d]$. A standard technique is that taking the first $|w|+r$ terms of the sum, we can approximate $d(w)$ to precision $2^{-r}$ in time $O((|w|+r) \cdot \max\{|w|+r, |w|^{2c+1}\})$. Such an $s$-gale can be defined for every set in $\mathcal{RL}_{\text{wk}}(2^{cn}, 2^{cn}, o(2^n))$. These gales are all computable within the same time bound, so we can apply a union lemma [19] to conclude that $\mathcal{RL}_{\text{wk}}(2^{cn}, 2^{cn}, o(2^n))$ has p-dimension at most $s$. □

## 4 Disjunctive and Conjunctive Reductions

In this section, as a warmup to our main theorem, we present two basic applications of Theorem 3.1. First, we consider disjunctive reductions.

**Theorem 4.1.** $P_d(\text{DENSE}^c)$ *has p-dimension 0.*

*Proof.* We will show that $P_d(\text{DENSE}^c) \subseteq \mathcal{RL}_{\text{wk}}(2^{2n}, 2^{2n}, o(2^n))$. For this, let $A \in P_d(\text{DENSE}^c)$ be arbitrary. Then there is a set $S \in \text{DENSE}^c$ and a reduction $f : \{0,1\}^* \to \mathcal{P}(\{0,1\}^*)$ computable in polynomial time $p(n)$ such that for all $x \in \{0,1\}^*$, $x \in A$ if and only if $f(x) \cap S \neq \emptyset$. Note that on an input of length $n$, all queries of $f$ have length bounded by $p(n)$. Also, since $S$ is nondense, for any $\epsilon > 0$ there are infinitely many $n$ such that

$$|S_{\leq p(n)}| \leq 2^{n^\epsilon}. \tag{4.1}$$

Let $Q_n = \bigcup_{|x| \leq n} f(x)$ be the set of all queries made by $f$ up through length $n$. Then $|Q_n| \leq 2^{n+1} p(n)$. Enumerate $Q_n$ as $q_1, \ldots, q_N$. Then each subset of $R \subseteq Q_n$ can be identified with its characteristic string $\chi_R \in \{0,1\}^N$ according to this enumeration. We define $\mathcal{C}_n$ to be the concept class of all monotone disjunctions on $\{0,1\}^N$ that have at most $2^{n^\epsilon}$ literals. Our target disjunction is

$$\phi_n = \bigvee_{i : q_i \in S} q_i,$$

which is a member of $\mathcal{C}_n$ whenever (4.1) holds. For any $x \in \{0,1\}^{\leq n}$,

$$x \in A \iff \phi_n(\chi_{f(x)}) = 1.$$

Given $x$, $\chi_{f(x)}$ can be computed in $O(2^{2n})$ time. Therefore $A \leq_{wk}^{O(2^{2n})} \mathcal{C} = (\mathcal{C}_n \mid n \in \mathbb{N})$. Since Winnow learns $\mathcal{C}_n$ making at most $2 \cdot 2^{n^\epsilon} \log |Q_n| + 2 = o(2^n)$ mistakes, it follows that $A \in \mathcal{RL}_{wk}(2^{2n}, 2^{2n}, o(2^n))$. □

Next, we consider conjunctive reductions.

**Theorem 4.2.** $P_c(\text{DENSE}^c)$ *has p-dimension 0.*

*Proof.* We will show that $P_c(\text{DENSE}^c) \subseteq \mathcal{RL}_{wk}(2^n, 2^{2n}, o(2^n))$. For this, let $A \leq_c^P S \in \text{DENSE}^c$. Then there is a reduction $f : \{0,1\}^* \to \mathcal{P}(\{0,1\}^*)$ computable in polynomial time $p(n)$ such that for all $x \in \{0,1\}^*$, $x \in A$ if and only if $f(x) \subseteq S$.

Fix an input length $n$, and let $Q_n = \bigcup_{|x| \leq n} f(x)$. Let $\epsilon > 0$ and consider the concept class

$$\mathcal{C}_n = \{\mathcal{P}(X) \mid X \subseteq Q_n \text{ and } |X| \leq 2^{n^\epsilon}\}.$$

Our target concept is

$$C_n = \mathcal{P}(S \cap Q_n).$$

For infinitely many $n$, $|S \cap Q_n| \leq |S_{\leq p(n)}| \leq 2^{n^\epsilon}$, in which case $C_n \in \mathcal{C}_n$. For any $x \in \{0,1\}^{\leq n}$, we have

$$x \in A \iff f(x) \in C_n.$$

Therefore $A \leq_{wk}^{p(n)} \mathcal{C} = (\mathcal{C}_n \mid n \in \mathbb{N})$.

The class $\mathcal{C}_n$ can be learned by a simple algorithm that makes at most $|X|$ mistakes when learning $\mathcal{P}(X)$. The hypothesis for $X$ is simply the union of all positive examples seen so far. More explicitly, the algorithm begins with the hypothesis $H = \emptyset$. In any stage, given an example $Q$, the algorithm predicts 'yes' if $Q \subseteq H$ and 'no' otherwise. If the prediction is 'no,' but $Q$ is revealed to be a positive example, then the hypothesis is updated as $H := H \cup Q$. The algorithm will never make a mistake on a negative example, and can make at most $|X|$ mistakes on positive examples.

This algorithm shows that $\mathcal{C} \in \mathcal{L}(2^{2n}, o(2^n))$, so $A \in \mathcal{RL}_{wk}(p(n), 2^{2n}, o(2^n))$. It follows that $P_c(\text{DENSE}^c) \subseteq \mathcal{RL}_{wk}(2^n, 2^{2n}, 2^{n^\epsilon})$. □

Since $\dim_p(E) = 1$, we have new proofs of the following results of Watanabe.

**Corollary 4.3.** (Watanabe [27]) $E \not\subseteq P_d(\text{DENSE}^c)$ *and* $E \not\subseteq P_c(\text{DENSE}^c)$. *That is, every $\leq_d^P$-hard or $\leq_c^P$-hard set for* $E$ *is dense.*

# 5  Adaptive Reductions

In this section we prove our main theorem, which concerns adaptive reductions that make a sublinear number of queries to a nondense set. It turns out that this problem can also be reduced to learning disjunctions.

In a surprising result (refuting a conjecture of Ko [14]), Allender, Hemaspaandra, Ogiwara, and Watanabe [1] showed that $P_{btt}(\text{SPARSE}) \subseteq P_d(\text{SPARSE})$. The disjunctive reduction they obtain will not be polynomial-time computable if the original reduction has more than a constant number of queries. However, in the proof of the following theorem we are still able to exploit their technique, and obtain an exponential-time reduction to a disjunction. Then we can apply the Winnow algorithm as in the previous section.

**Theorem 5.1.** *For all $\alpha < 1$, $P_{n^\alpha-T}(\text{DENSE}^c)$ has p-dimension 0.*

*Proof.* Let $L \leq^p_{n^\alpha-T} S \in \text{DENSE}^c$ via some oracle machine $M$. We will show how to reduce $L$ to a class of disjunctions.

Fix an input length $n$. For an input $x \in \{0,1\}^{\leq n}$, consider using each $z \in \{0,1\}^{n^\alpha}$ as the sequence of yes/no answers to $M$'s queries. Each $z$ causes $M$ to produce a sequence of queries $w_0^{x,z}, \ldots, w_{k(x,z)}^{x,z}$, where $k(x,z) < n^\alpha$, and an accepting or rejecting decision. Let $Z_x \subseteq \{0,1\}^{n^\alpha}$ be the set of all query answer sequences that cause $M$ to accept $x$. Then we have $x \in L$ if and only if

$$(\exists z \in Z_x)(\forall 0 \leq j \leq k(x,z)) \; S[w_j^{x,z}] = z[j],$$

which is equivalent to

$$(\exists z \in Z_x)(\forall 0 \leq j \leq k(x,z)) \; z[j] \cdot w_j^{x,z} \in S^c \oplus S,$$

where $S^c \oplus S$ is the disjoint union $\{0x \mid x \in S^c\} \cup \{1x \mid x \in S\}$.

A key part of the proof that $P_{btt}(\text{SPARSE}) \subseteq P_d(\text{SPARSE})$ in [1] is to show that $P_{1-tt}(\text{SPARSE})$ is contained in $P_d(\text{SPARSE})$. The same argument yields that

$$P_{1-tt}(\text{DENSE}^c) \subseteq P_d(\text{DENSE}^c).$$

Therefore, there is a set $U \in \text{DENSE}^c$ such that $S^c \oplus S \leq^p_d U$. Letting $g$ be this polynomial-time disjunctive reduction, we have $x \in L$ if and only if

$$(\exists z \in Z_x)(\forall 0 \leq j \leq k(x,z)) \; g(z[j] \cdot w_j^{x,z}) \cap U \neq \emptyset.$$

For each $z \in Z_x$, let

$$H_{x,z} = \{\langle u_0, \ldots, u_{k(x,z)} \rangle \mid (\forall 0 \leq j \leq k(x,z)) \; u_j \in g(z[j] \cdot w_j^{x,z})\}.$$

Define

$$A_n = \{\langle u_0, \ldots, u_k \rangle \mid k < n^\alpha \text{ and } (\forall 0 \leq j \leq k) \; u_j \in U\}.$$

Then we have $x \in L$ if and only if

$$(\exists z \in Z_x)(\exists v \in H_{x,z}) \; v \in A_n.$$

Letting
$$H_x = \bigcup_{z \in Z_x} H_{x,z},$$
we can rewrite this as
$$x \in L \iff H_x \cap A_n \neq \emptyset. \tag{5.1}$$

Let $r(n)$ be a polynomial bounding the number of queries $g$ outputs on an input of form $z[j] \cdot w_j^{x,z}$, where $|x| \leq n$. Then $|H_{x,z}| \leq r(n)^{n^\alpha}$, so
$$|H_x| \leq |Z_x| \cdot r(n)^{n^\alpha} \leq 2^{n^\alpha \cdot (1 + \log r(n))}. \tag{5.2}$$

Also,
$$|A_n| \leq n^\alpha \cdot |U_{\leq r(n)}|^{n^\alpha}.$$

Let $\epsilon \in (0, 1 - \alpha)$, and let $\delta \in (\alpha + \epsilon, 1)$. Then since $U$ is nondense, for infinitely many $n$, we have $|U_{\leq r(n)}| \leq 2^{n^\epsilon}$. This implies
$$(\exists^\infty n)\ |A_n| \leq n^\alpha \cdot 2^{n^{\alpha+\epsilon}} \leq 2^{n^\delta}. \tag{5.3}$$

Let
$$H_n = \bigcup_{x \in \{0,1\}^{\leq n}} H_x.$$

Then from (5.2), $|H_n| \leq 2^{2n}$ if $n$ is sufficiently large.

Enumerate $H_n$ as $h_1, \cdots, h_N$. We identify any $R \subseteq H_n$ with its characteristic string $\chi_R^{(n)} \in \{0,1\}^N$ according to this enumeration. Let $\mathcal{C}_n$ be the concept class of all monotone disjunctions on $\{0,1\}^N$ that have at most $2^{n^\delta}$ literals.

Define the disjunction
$$\phi_n = \bigvee_{i:h_i \in A_n} h_i,$$

$\phi_n = \bigvee_{i:h_i \in A_n} h_i$, which by (5.3) is in $\mathcal{C}_n$ for infinitely many $n$. For any $x \in \{0,1\}^{\leq n}$, from (5.1) it follows that
$$x \in L \iff \phi_n(\chi_{H_x}^{(n)}) = 1.$$

Given $x \in \{0,1\}^{\leq n}$, we can compute $\chi_{H_x}^{(n)}$ in $O(2^n \cdot \text{poly}(n) + |H_n|)$ time. Therefore, letting $\mathcal{C} = (\mathcal{C}_n \mid n \in \mathbb{N})$, we have $L \leq_{\text{wk}}^{2^{2n}} \mathcal{C}$. Since $\mathcal{C}_n$ is learnable by Winnow with at most $2 \cdot 2^{n^\delta} \cdot \log |H_n| + 2 = o(2^n)$ mistakes, it follows that $L \in \mathcal{RL}_{\text{wk}}(2^{2n}, 2^{2n}, o(2^n))$. □

As a corollary, we have a positive answer to the question of Lutz and Mayordomo [21] mentioned in the introduction:

**Corollary 5.2.** *For all $\alpha < 1$, $\text{P}_{n^\alpha - \text{T}}(\text{DENSE}^c)$ has p-measure 0.*

**Corollary 5.3.** *For all $\alpha < 1$, $\text{E} \not\subseteq \text{P}_{n^\alpha - \text{T}}(\text{DENSE}^c)$. That is, every $\leq_{n^\alpha - \text{T}}^{\text{P}}$-hard set for E is dense.*

If we scale down from nondense sets to sparse sets, the same proof technique can handle more queries.

**Theorem 5.4.** $P_{o(n/\log n)-T}(\text{SPARSE})$ *has strong p-dimension 0.*

*Proof.* Let $L \leq^p_{f(n)-T} S \in \text{SPARSE}$ via some oracle machine $M$, where $f(n) = o(n/\log n)$.

Fix an input length $n$. For an input $x \in \{0,1\}^n$, each query answer sequence $z \in \{0,1\}^{f(n)}$ causes $M$ to produce a sequence of queries $w_0^{x,z}, \ldots, w_{k(x,z)}^{x,z}$, where $k(x,z) < f(n)$, and an accepting or rejecting decision. Let $Z_x \subseteq \{0,1\}^{f(n)}$ be the set of all query answer sequences that cause $M$ to accept $x$. Then we have $x \in L$ if and only if

$$(\exists z \in Z_x)(\forall 0 \leq j \leq k(x,z)) \; z[j] \cdot w_j^{x,z} \in S^c \oplus S.$$

Since $P_{1-tt}(\text{SPARSE}) \subseteq P_d(\text{SPARSE})$, there is a set $U \in \text{SPARSE}$ such that $S^c \oplus S \leq^p_d U$. Letting $g$ be this polynomial-time disjunctive reduction, we have $x \in L$ if and only if

$$(\exists z \in Z_x)(\forall 0 \leq j \leq k(x,z)) \; g(z[j] \cdot w_j^{x,z}) \cap U \neq \emptyset.$$

As before, we can define sets $H_x$ and $A_n$ so that

$$x \in L \iff H_x \cap A_n \neq \emptyset.$$

Let $r(n)$ be a polynomial bounding the number of queries $g$ outputs on an input of form $z[j] \cdot w_j^{x,z}$, where $|x| = n$. Then

$$|H_x| \leq |Z_x| \cdot r(n)^{f(n)} \leq 2^{f(n) \cdot (1 + \log r(n))},$$

so we have $|H_x| \leq 2^n$ if $n$ is sufficiently large because $f(n) = o(n/\log n)$. Letting $H_n = \bigcup_{x \in \{0,1\}^n} H_x$, we have $|H_n| \leq 2^{2n}$.

Also,

$$|A_n| \leq f(n) \cdot |U_{\leq r(n)}|^{f(n)}.$$

Let $q(n)$ be a polynomial such that $|U_{\leq r(n)}| \leq q(n)$ for all $n$. Then

$$|A_n| \leq f(n) \cdot q(n)^{f(n)} \leq 2^{f(n) \log q(n) + \log f(n)}.$$

Let $v(n) = f(n) \log q(n) + \log f(n)$. Notice that $v(n) = o(n)$ because $f(n) = o(n/\log n)$.

As before, we enumerate $H_n$ as $h_1, \cdots, h_N$ and identify any $R \subseteq H_n$ with its characteristic string $\chi_R^{(n)} \in \{0,1\}^N$. Let $\mathcal{C}_n$ be the concept class of all monotone disjunctions on $\{0,1\}^N$ that have at most $2^{v(n)}$ literals. The disjunction $\phi_n = \bigvee_{i: h_i \in A_n} h_i$, is in $\mathcal{C}_n$ for every $n$. For any $x \in \{0,1\}^n$, we have

$$x \in L \iff \phi_n(\chi_{H_x}^{(n)}) = 1.$$

Given $x \in \{0,1\}^n$, we can compute $\chi_{H_x}^{(n)}$ in $O(2^n \cdot \text{poly}(n) + |H_n|)$ time. Therefore, letting $\mathcal{C} = (\mathcal{C}_n \mid n \in \mathbb{N})$, we have $L \leq^{2^{2n}}_{\text{str}} \mathcal{C}$. Since $\mathcal{C}_n$ is learnable by Winnow with at most $2 \cdot 2^{v(n)} \cdot \log |H_n| + 2 = o(2^n)$ mistakes, it follows that $L \in \mathcal{RL}_{\text{str}}(2^{2n}, 2^{2n}, o(2^n))$. □

The following corollary improves the result of Fu [8] that $E \not\subseteq P_{o(n/\log n)-T}(\text{TALLY})$.

**Corollary 5.5.** $E \not\subseteq P_{o(n/\log n)-T}(\text{SPARSE})$.

Since Wilson constructed an oracle relative to which $E \subseteq P_{O(n)-tt}(SPARSE)$ [28, 20], Corollary 5.5 is near the limits of relativizable techniques.

In Theorem 5.4, we used strong dimension, which raises a technical point. The results about reductions to $DENSE^c$ cannot be strengthened to strong p-dimension simply because the class $DENSE^c$ itself has strong dimension 1. This is because being nondense is an infinitely-often property [9]. However, if we replace $DENSE^c$ by SPARSE in any of our results, the proofs can be adapted to show that the resulting class has strong p-dimension 0. We can also obtain strong dimension results by substituting the larger class $DENSE_{i.o.}^c$, where $DENSE_{i.o.}$ is the class of all $L$ that satisfy $(\exists \epsilon > 0)(\exists^\infty n) \, |L_{\leq n}| > 2^{n^\epsilon}$.

## 6 Hard Sets for NP

The hypothesis "NP has positive p-dimension," written $\dim_p(NP) > 0$, was first used in [11] to study the inapproximability of MAX3SAT. This positive dimension hypothesis is apparently much weaker than Lutz's often-investigated $\mu_p(NP) \neq 0$ hypothesis, but is a stronger assumption than $P \neq NP$:

$$\mu_p(NP) \neq 0 \Rightarrow \dim_p(NP) = 1 \Rightarrow \dim_p(NP) > 0 \Rightarrow P \neq NP.$$

The measure hypothesis $\mu_p(NP) \neq 0$ has many plausible consequences that are not known to follow from $P \neq NP$ (see e.g. [21]). So far few consequences of $\dim_p(NP) > 0$ are known. The following corollary of our results begins to remedy this.

**Theorem 6.1.** *If* $\dim_p(NP) > 0$, *then every set that is hard for* NP *under* $\leq_d^P$*-reductions,* $\leq_c^P$*-reductions, or* $\leq_{n^\alpha-T}^P$*-reductions ($\alpha < 1$) is dense, and every set that is hard under* $\leq_{o(n/\log n)-T}^P$*-reductions is not sparse.*

The consequences in Theorem 6.1 are much stronger than what is known to follow from $P \neq NP$. If $P \neq NP$, then no $\leq_{btt}^P$-hard or $\leq_c^P$-hard set is sparse [26, 3], but it is not known whether hard sets under disjunctive reductions or unbounded Turing reductions can be sparse.

Another result is that if $NP \neq RP$, then no $\leq_d^P$-hard set for NP is sparse [7, 6]. It is interesting to see that while the hypotheses $\dim_p(NP) > 0$ and $NP \neq RP$ are apparently incomparable, they both have implications for the density of the disjunctively-hard sets for NP.

## 7 Conclusion

Our connection between online learning and resource-bounded dimension appears to be a powerful tool for computational complexity. We have used it to give relatively simple proofs and improvements of several previous results.

An interesting observation is that for all reductions $\leq_\tau^P$ for which we know how to prove "every $\leq_\tau^P$-hard set for E is dense," by the results presented here we can actually prove "$P_\tau(DENSE^c)$ has p-dimension 0." Indeed, we have proven the strongest results for Turing reductions in this way.